\documentclass{article}

\usepackage[english]{babel}

\usepackage[letterpaper,top=2cm,bottom=2cm,left=3cm,right=3cm,marginparwidth=1.75cm]{geometry}

\usepackage{amsmath,amssymb}
\usepackage{graphicx}
\usepackage[rightcaption]{sidecap} 
\usepackage{subcaption}

\usepackage{siunitx}
\usepackage[colorlinks=true, allcolors=blue]{hyperref}
\usepackage{xcolor}
\usepackage{xspace}
\usepackage{qcircuit}

\usepackage{tikz}
\usetikzlibrary{shapes, arrows.meta, positioning}
\tikzstyle{startstop} = [rectangle, rounded rectangle, 
minimum width=3cm, 
minimum height=1cm,
text centered, 
draw=black]

\tikzstyle{process} = [rectangle, 
minimum width=2cm, 
minimum height=1cm, 
text centered, 
text width=3cm, 
draw=black]

\tikzstyle{data} = [trapezium, 
trapezium left angle=60, 
trapezium right angle=120, 
minimum width=2cm, 
minimum height=1cm, 
text width=2.5cm, 
text centered, 
draw=black]

\tikzstyle{decision} = [diamond, 
text width=2cm,
minimum width=2cm, 
minimum height=1cm, 
aspect=1.5, 
text width=1cm, 
text centered, 
draw=black]
\tikzstyle{arrow} = [thick,->,>=stealth]

\newcommand{\autoCAS}{\textsc{autoCAS}\xspace}
\newcommand{\AutoRXN}{AutoRXN\xspace}

\newcommand{\pauli}[1]{\mathrm{#1}}
\newcommand{\pauliI}{\pauli{I}}
\newcommand{\pauliX}{\pauli{X}}
\newcommand{\pauliY}{\pauli{Y}}
\newcommand{\pauliZ}{\pauli{Z}}
\newcommand{\pauliIX}{\pauli{IX}}
\newcommand{\pauliXI}{\pauli{XI}}
\newcommand{\pauliIZ}{\pauli{IZ}}
\newcommand{\pauliZI}{\pauli{ZI}}
\newcommand{\pauliXX}{\pauli{XX}}
\newcommand{\pauliXZ}{\pauli{XZ}}
\newcommand{\pauliZX}{\pauli{ZX}}
\newcommand{\pauliZZ}{\pauli{ZZ}}
\newcommand{\logicalpauliXX}{\bar{\pauli{X}}\bar{\pauli{X}}}
\newcommand{\logicalpauliXZ}{\bar{\pauli{X}}\bar{\pauli{Z}}}
\newcommand{\logicalpauliZX}{\bar{\pauli{Z}}\bar{\pauli{X}}}
\newcommand{\logicalpauliZZ}{\bar{\pauli{Z}}\bar{\pauli{Z}}}

\newcommand{\RX}{\mathrm{R}_\pauli{X}}
\newcommand{\RY}{\mathrm{R}_\pauli{Y}}
\newcommand{\gateS}{\pauli{S}}
\newcommand{\expectation}[1]{\langle #1 \rangle}
\newcommand{\ket}[1]{|#1\rangle}
\newcommand{\bra}[1]{\langle #1 |}
\newcommand{\gs}{\Psi_\mathrm{gs}}
\newcommand{\outputstate}{\Phi(\alpha,\beta,\gamma)}

\newcommand{\RR}{\mathbb{R}}
\DeclareMathOperator{\Tr}{Tr}
\newcommand{\Cfour}{$C_4$\xspace}

\title{%
{End-to-End Quantum Simulation of a Chemical System}}
\author{Microsoft Azure Quantum\footnote{(in alphabetical order): 
Wim van Dam, Hongbin Liu, Guang Hao Low, Adam Paetznick, Andres Paz, Marcus Silva, Aarthi Sundaram, Krysta Svore, Matthias Troyer.
Work by GHL for this paper was done while at Microsoft.}
}

\begin{document}
\maketitle

\begin{abstract}
We demonstrate the first end-to-end integration of high-performance computing (HPC), reliable quantum computing, and AI in a case study on catalytic reactions producing chiral molecules. 
We present a hybrid computation workflow to determine the strongly correlated reaction configurations and estimate, for one such configuration, its active site's ground state energy. 
We combine 1) the use of HPC tools like AutoRXN and AutoCAS to systematically identify the strongly correlated chemistry within a large chemical space with 2) the use of logical qubits in the quantum computing stage to prepare the quantum ground state of the strongly correlated active site, demonstrating the advantage of logical qubits compared to physical qubits, and 3) the use of optimized quantum measurements of the logical qubits with so-called classical shadows to accurately predict various properties of the ground state including energies. 
The combination of HPC, reliable quantum computing, and AI in this demonstration serves as a proof of principle of how future hybrid chemistry applications will require integration of large-scale quantum computers with classical computing to be able to provide a measurable quantum advantage.
\end{abstract}

\newpage


\section{Introduction}

Quantum computing is poised to dramatically change today's computing paradigm due to its ability to leverage quantum mechanical phenomena such as superposition and entanglement. 
Chemistry stands as one of the most promising areas for quantum computing applications, as the behavior of electrons within molecules, particularly during chemical reactions, is inherently quantum mechanical. 
These reactions, which involve complex bond formations and breakages, are driven by strongly correlated electron interactions~\cite{feynman2018simulating,hoefler2023disentangling,liu2022prospects}. 
Consequently, understanding and simulating chemical reactions requires a deep quantum mechanical treatment.

While it is tempting to view chemical reactions as a series of discrete steps, they often involve a vast network of interconnected reaction pathways~\cite{unsleber2020exploration}. 
Exploring and simulating the entire reaction network, rather than a single reaction, is critical to understanding the full mechanistic picture. 
However, traditional computational approaches, while effective for many simpler cases, struggle with the scale and complexity of accurately mapping these networks. 
In classical computational chemistry, approximate methods such as Density Functional Theory (DFT)~\cite{HKDFT,KSDFT} have proven efficient and widely adopted. 
Despite their success, these methods do not reach the accuracy required for highly correlated systems where electron-electron interactions become critical, such as during bond breaking and formation. 
Here, more accurate methods like Full Configuration Interaction (FCI)~\cite{FCI} are required, as they provide exact solutions to the electronic Schr{\"o}dinger equation by considering all possible electron configurations. Unfortunately, FCI scales exponentially with the size of the system, making it computationally intractable for anything beyond very small molecules on classical computers. To address these limitations, the concept of active spaces has been developed. Active space methods, such as Complete Active Space Configuration Interaction (CAS-CI)~\cite{CAS}, focus computational resources on the most important, or ``active'', molecular orbitals where significant electron correlation occur, while treating the remaining orbitals more approximately. However, even active space methods, though significantly reducing the computational cost compared to FCI, face their own limitations. For many realistic chemical systems, the active spaces become too large, rendering CAS-CI impractical on classical hardware. This is where quantum computing offers a new frontier. Quantum computers, which naturally handle superposition and entanglement, hold the promise of solving some active space configurations currently impossible for classical machines. 

While to solve large active spaces, quantum computers are expected to require a large number of qubits~\cite{von2021quantum,qre2022}, there has been a plethora of demonstration experiments on small, albeit still classically tractable, systems over the past decade. Early experiments analyzed molecular hydrogen and ran variants of quantum phase estimation (QPE) on silicon chips~\cite{lanyon2010towards, paesani2017experimental}. 
These experiments achieved chemical accuracy using hardware specific operations and gates for their demonstration. 
Other examples for larger systems include computing the bond energy for Helium Hydride ion in a solid-state quantum spin register using phase estimation~\cite{wang2015quantum}. Over the years, there have been several demonstrations of the variational quantum eigensolver (VQE) algorithm on noisy intermediate-scale quantum (NISQ) systems, ranging from $2$ to $10{+}$ physical qubits. A comprehensive list of recent results and techniques can be found in~\cite{fedorov2022vqe, tilly2022variational}. 
On superconducting qubits, an initial comparison in Ref.~\cite{o2016scalable} between VQE and QPE showed that the former has shorter circuits and also provides estimates to within chemical accuracy for specific bond lengths. 
In comparison, QPE was found to be unable to match that precision on NISQ systems -- requiring deeper and more noise-resilient circuits to enable a reliable solution. Additionally, by changing the parameters in the VQE circuit, it was possible to map out the full potential energy surface for $\text{H}_2$ for all bond lengths. 

Beyond the molecular toy systems that are of scientific interest, Ref.~\cite{nutzel2024solving} performs 10-qubit VQE on unencoded trapped-ion qubits on an industrially relevant chemical problem. They compute the ground state energy for a strongly correlated metal chelate up to chemical accuracy using a Hartree-Fock state to start with, and perform more detailed post-processing to generate energy estimates.

More recently, an error-detected version of the $\text{H}_2$ experiment in Ref.~\cite{urbanek2020error} on superconducting qubits, using a $[[4, 2, 2]]$ code, showed better performance than the corresponding unencoded version on the same device while mapping out the potential energy surface. 
Building on this is the recent Ref.~\cite{gowrishankar2024} investigated the noise sensitivity of a $[[4, 2, 2]]$ encoded algorithm for this
task.
An error-detected Bayesian quantum phase estimation experiment on $\text{H}_2$ was performed on Quantinuum's H1 system in Ref.~\cite{Yamamoto2024} using a $[[6, 4, 2]]$ code. While the energy estimate obtained using the logical-qubit computation was an order of magnitude better than the estimate produced through unencoded qubits, this result also demonstrates how encoding with logical qubits can allow for relatively deep circuits with logical depth $> 100$ to be run effectively and reliably on hardware.

Despite the progress of simulating chemical systems on quantum computers, there remain significant challenges to scaling up the size of simulable system possible on quantum computers. First, most chemical systems that have been studied are either artificial (e.g., H4, stretched LiH) or randomly picked. Furthermore, there is a lack of systematic approach to finding and identifying the molecules that would benefit from study by quantum computers, for example, those that may not be solvable or well-approximated classically or possess certain characteristics. 
Second, the noise of physical qubits, no matter the type, introduces errors in the total energy that diminish the value of a more exact solution with a quantum computer.  That is, using physical qubits in a quantum computation to determine the energy is simply too noisy as the chemical system scales up; reliable logical qubits will be required to combat the physical noise of the quantum computer. 
Lastly, scientists are interested in many properties other than energies in chemistry. However, most of the quantum experiments so far focuses solely on energy. 

While there have been papers reporting the implementation and demonstration of  individual components of the hybrid workflow presented herein, such as HPC calculations, or quantum simulations, or classical shadows, in this work we present and demonstrate the first end-to-end hybrid workflow that combines HPC, logical quantum computing, and AI on a real chemistry problem. 
We use our hybrid workflow to study a catalysis problem focused on the chiral reaction of acetophenone, elucidating the reaction mechanism with HPC, simulating the quantum behavior of the active space with error-detecting quantum circuits, and finally using AI techniques to prepare and sample the ground state energies. This hybrid workflow provides a glimpse into how these technologies can work together to ultimately solve more complex chemical problems with unprecedented accuracy as reliable quantum computers scale.

\section{The hybrid workflow and  chemistry problem \label{section:chemistry background}}
\begin{SCfigure}[1.6]
\begin{tikzpicture}[node distance = 1.5cm]
\node (start) [startstop] {start}; 
\node (hpc) [process, below of=start] {classical HPC calculations};
\node (hpcdata) [data,below of=hpc] {active space Hamiltonian};
\node (qc) [process,below of=hpcdata] {reliable quantum computing (QC)};
\node (qcdata) [data,below of=qc] {classical QC data};
\node (ai) [process,below of=qcdata] {classical AI post-processing};
\node (aidata) [data,below of=ai] {classical model and predictions};
\node (finish) [startstop,below of=aidata] {finish};

\draw [arrow] (start) -- (hpc);
\draw [arrow] (hpc) -- (hpcdata);
\draw [arrow] (hpcdata) -- (qc);
\draw [arrow] (qc) -- (qcdata);
\draw [arrow] (qcdata) -- (ai);
\draw [arrow] (ai) -- (aidata);
\draw [arrow] (aidata) -- (finish);
\end{tikzpicture}
\caption{%
Flowchart of our overall approach combining high-performance computing (HPC), reliable quantum computing (QC), and artificial intelligence (AI) to calculate properties of ground states in quantum chemistry.  
Classical HPC is used to find the candidate molecules that have the most need to be evaluated by quantum computation and to reduce the dimension of the relevant active space such that it can be analyzed with quantum computation. 
The outcome of this HPC calculation (``active space Hamiltonian'') is used to construct a reliable quantum circuit for the wave function preparation and for defining the measurement bases that are used to sample the quantum state.   
Using AI techniques the quantum measurement data is used to define a succinct classical model of the exponentially large quantum state that predicts, with high precision, the relevant quantities, such as the ground state energy.
\label{figure:flowchart_hpc_qc_ai}}
\end{SCfigure}
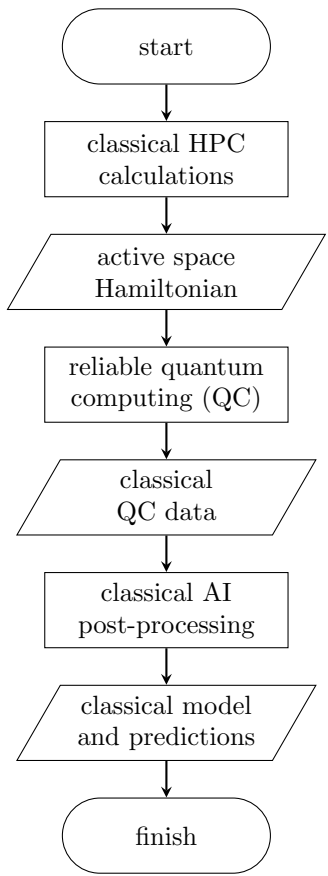

Our overall computation workflow is summarized in Figure~\ref{figure:flowchart_hpc_qc_ai}. We use a combination of cloud HPC, reliable quantum computing, and AI to address the three aforementioned challenges. 
Cloud HPC simulations are used to identify the specific structures of a given chemical problem that are strongly correlated. Quantum computation with encoded qubits and operations are used to accurately measure the correlation of the specific structures. 
Finally, AI models are used to generalize the quantum measurement for more broad, accurate property predictions.  
\sidecaptionvpos{figure}{c}

\subsection{The chemistry problem: catalysis and chirality}
Chirality plays a critical role in chemistry and biology, particularly in drug synthesis, where the difference between two enantiomers (mirror-image molecules) can mean the difference between therapeutic benefit and harmful effects.
Therefore, reactions to produce specific enantiomers are critical for biological functionalities and are often used in drug and agrichemical industries.
Despite the progress in the field~\cite{Nobel2001}, there is still great desire to design new catalysts that improve the selectivity of the chiral reactions, as well as more  sustainable and cheaper catalysts. 
Computation is the only feasible way to validate huge amounts of novel and hypothetical catalysts.  
However, chiral reactions are extremely challenging to compute as they require very precise estimation of the reaction energy barrier of each enantiomer. 
Even small errors (as small as \num{1} kcal/mol) in the barrier prediction change could lead to a significantly different selectivity and thus wrong evaluations. 
In this work, we focus on a chiral reaction involving acetophenone and a P-N-N-P iron catalyst. Understanding the energy barriers in such reactions is essential for improving enantiomer selectivity, but achieving the required precision is in general beyond classical computational methods.

\subsection{Identifying the reaction network and strongly correlated structures}
\begin{figure}
\begin{center}
\includegraphics[width=0.6\textwidth]{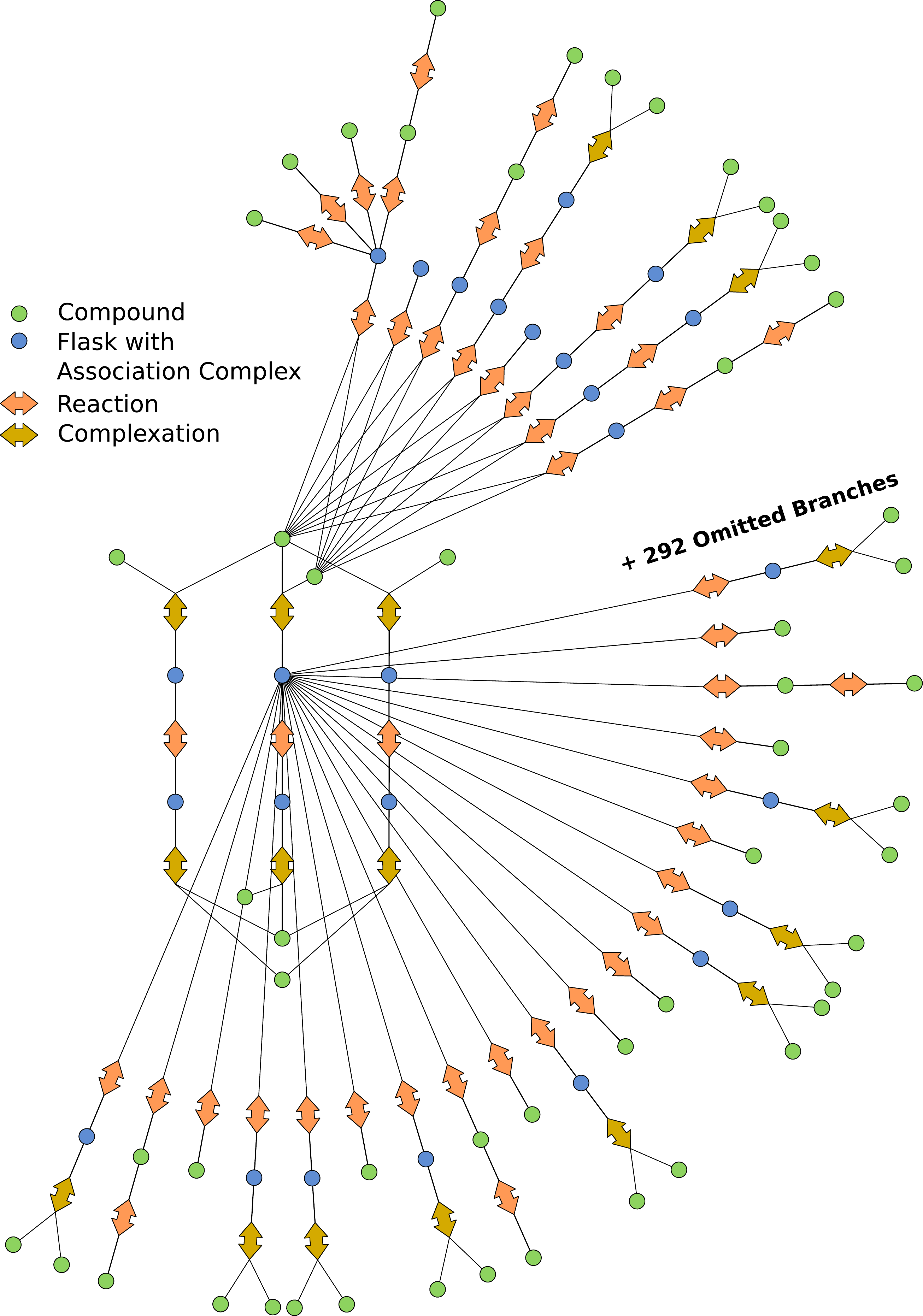} 
\end{center}
\caption{%
The full reaction network explored from \AutoRXN for the P-N-N-P Iron catalytic reactions. 
The center loop represents the main reaction pathways to the chiral products, while all extending branches represent the side reactions.  
\label{figure:reaction_network}}
\end{figure}

To begin, we elucidate the reaction network using cloud HPC, specifically using tools available through the Azure Quantum Elements platform~\cite{AQE}.
On the Azure Quantum Elements platform~\cite{AQE,Liu}, we used \AutoRXN~\cite{AutoRXN} to map out the reaction network of the P-N-N-P iron catalyst in the first-ever automated reaction network analysis in the cloud~\cite{Liu,AutoRXN}.  
More than one million DFT gradient calculations are conducted to generate \num{2227} elementary steps and \num{3150} unique configurations in the reaction network.
\AutoRXN identifies the reaction pathways to the products with both left and right-hand chirality, as well as many possible side reactions and byproducts (see Figure~\ref{figure:reaction_network}). 
However, the approximations in DFT introduce errors to the calculations and the errors could be significant in certain cases, as some of the structures within the reaction network express strong quantum correlations. 
For those cases, a much more accurate energy evaluation is needed. 
As an alternative to FCI, quantum computation promises to be extremely beneficial in these scenarios. 

\begin{figure}
\includegraphics[width=\textwidth]{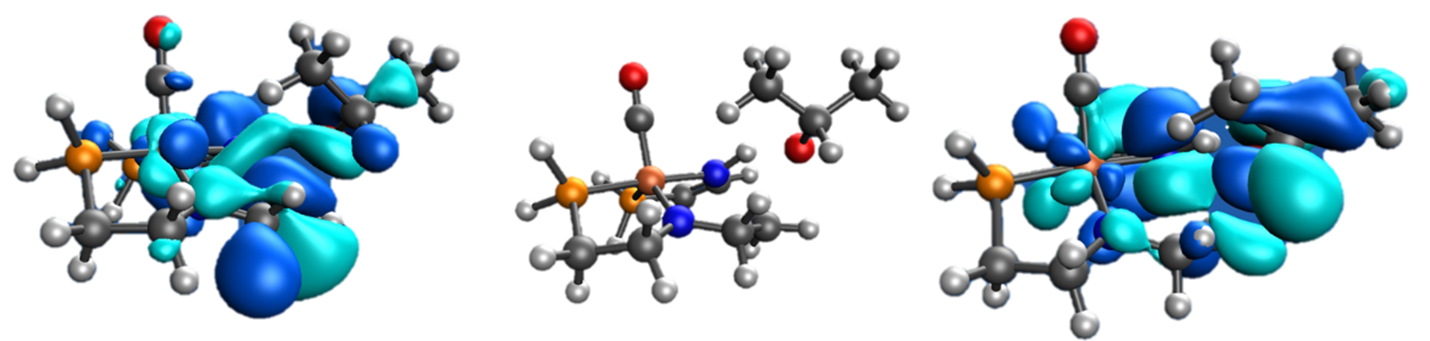} 
\caption{An example of a correlated molecule (middle), and the orbitals containing the entangled electrons (left, right) in the active space.  
\label{figure:active_space}}
\end{figure}

One of the critical steps in estimating the precise energy barriers of a reaction network is to identify those intermediates that are  strongly correlated, meaning that entanglement between electrons plays a significant role. 
We utilized the coupled-cluster workflow on Azure Quantum Elements to initially screen for strongly correlated structures~\cite{AutoRXN}. 
Among all 3150 structures in the reaction network, five of them appear to show strong correlation based on the coupled-cluster diagnostics. 
All of those strongly correlated structures have been run through the \autoCAS~\cite{AutoCAS} workflow on  Azure Quantum Elements. \autoCAS essentially conducts the single orbital entropy diagnostics based on the density matrix renormalization group (DMRG) calculations to identify the correlated orbitals and extracts them into so-called active space (see Figure~\ref{figure:active_space}).  
The final outcome of the \autoCAS workflow is an FCIDump file that contains the one- and two-body Hamiltonian describing the active space.  

\subsection{Preparing the ground state of a strongly correlated molecule}
The five configurations that require high accuracy treatment can be found in  Figure 11 of Ref.~\cite{AutoRXN}. Their active space sizes range from 2 orbitals to 59 orbitals. Quantum computers, with enough scale and reliability, are desirable to solve for the large active spaces, however, in this work we choose the smallest active space to demonstrate a proof-of-concept for a future, more scaled workflow. 

The 2-orbital active space we pick is based on the fourth structure in Figure 11 of \cite{AutoRXN}. It corresponds to the transition state of hydrogen loading onto the catalyst.

The Hamiltonian of the active space in the second quantization form can be written as:
\begin{align}
\hat{H} = & \left( -1.172 \, a_1^\dagger a_1 + -0.031 \, a_1^\dagger a_2 + -0.031 \, a_2^\dagger a_1 + -1.054 \, a_2^\dagger a_2 \right) \\
& + \frac{1}{2} \Big( 0.303 \, a_1^\dagger a_1^\dagger a_1 a_1 + 0.002 \, a_1^\dagger a_2^\dagger a_1 a_1 + 0.175 \, a_1^\dagger a_1^\dagger a_2 a_2 \\
& \quad + 0.015 \, a_1^\dagger a_2^\dagger a_1 a_2 + 0.002 \, a_1^\dagger a_2^\dagger a_2 a_2 + 0.283 \, a_2^\dagger a_2^\dagger a_2 a_2 \Big)
\end{align}

We then encode the Hamiltonian in qubits: 
\begin{align}
H & = 
\num{- 1.99134} 
-\qty{0.02882925}{(\pauliX_1+\pauliX_2)}
+ \qty{0.0541175}{(\pauliZ_1+\pauliZ_2)} \nonumber \\
& \quad 
+ \qty{0.01495595}{\pauliX_1\pauliX_2}
+ \qty{0.000151287}{(\pauliX_1\pauliZ_2+\pauliZ_1\pauliX_2)}
+ \qty{0.05900925}{\pauliZ_1\pauliZ_2}, 
\label{eq:hamiltonian}
\end{align}
where the coefficients are in Hartree units. 

As quantum computers scale and become more reliable, QPE will always be favored for solving the Hamiltonian due to its advantages in error controls. In the near term, however, especially for small active spaces, VQE can also be used, given the wave function ansatz can still resemble the FCI wave function. For this demonstration herein, we calculate the exact ground state using classical methods
\begin{align}\label{eq:fciclassical}
\ket{\gs} & = 
\qty{0.070866}{\ket{00}} + 
\qty{0.499955}{\ket{01}} + 
\qty{0.499955}{\ket{10}} + 
\qty{0.703611}{\ket{11}}, 
\end{align}
and prepare the exact state on the quantum computer. 
The generalization to VQE is straightforward, and merely requires more runtime for additional measurements during each classical optimization step.

The quantum circuits of the ground state $\ket{\gs}\in \RR^4$ can be produced by the two-qubit logical circuit 
\begin{equation}\label{eq:logical_circuit}
\Qcircuit @C=1.4em @R=1.2em {
\lstick{\ket{0}}  & \gate{\RY(\alpha)}&\ctrl{1} & \qw                & \qw \\ 
\lstick{\ket{0}}  & \gate{\RY(\beta)} & \targ   & \gate{\RY(\gamma)} & \qw \\ 
}
\end{equation}
with three parameterized single-qubit $\pauliY$-rotations
\begin{align}
\RY(\theta) & = 
\begin{pmatrix} 
\cos(\theta/2) & -\sin(\theta/2) \\
\sin(\theta/2) & \cos(\theta/2)
\end{pmatrix},
\end{align}
and the appropriate angles $\alpha$, $\beta$, and $\gamma$. 
We denote the output state by $\outputstate$. 
Combining Equation~\ref{eq:fciclassical}, the ground state can be produced by the circuit of Equation~\ref{eq:logical_circuit} using the angles
\begin{align}
(\alpha, \beta, \gamma) & = (\num{2.08293}, \num{2.04776}, \num{0.81221}). \label{eq:optimal_angles}
\end{align}
In this work we focus on the precision of sampling the ground state accurately on the logical qubits, hence, we will  fix the angles $(\alpha,\beta,\gamma)$ and do not perform the parameter optimization, as done in other "traditional" VQE examples. 


\subsection{Classical Shadows}
Once the wave function is prepared as a quantum state, any chemical property of interest can be in principle estimated from it. 
We do not directly estimate the energy through measurements, but instead perform simpler, random Pauli measurements to learn the so-called classical shadows. 
This has two advantages: first, the measurements are easier than measuring the four point functions needed for  direct energy measurements, and second, it gives a model for the quantum state that can be used to estimate arbitrary expectation values.~\cite{Huang2020, Huang2022,low2022classical}


The method repeatedly applies the following procedure to the unknown state $\rho$: 
\begin{enumerate}
\item Apply a random unitary $U$ that rotates the state (i.e., $\rho \mapsto U \rho U^{\dagger}$).
\item Measure all qubits in the computational $\pauliZ$ basis, yielding the bit string $\hat{b}$ as the measurement outcome. 
At this point, the state $U^{\dagger} \ket{\hat{b}} \bra{\hat{b}} U$ serves as the description of the state the original $\rho$ has collapsed to.
When averaged over the choice of unitary and the measurement outcome, this random snapshot contains unbiased information about the input state $\rho$. 
Furthermore, the process of mapping $\rho$ to $U^{\dagger} \ket{b} \bra{b}U$ acts as a quantum channel $\mathcal{M}$ defined by 
\begin{align}
 \mathcal{M} (\rho) & = \mathbb{E} [U^{\dagger} \ket{{b}}\bra{{b}} U] \text{, such that~} \rho = \mathcal{M}^{-1} \mathbb{E} [U^{\dagger} \ket{{b}}\bra{b} U].
\end{align}
The inverted channel, $\mathcal{M}^{-1}$, may not be a physical quantum channel, however, it can be classically applied in post-processing to the measurement outcome $b$ to generate the random snapshot $\hat{\rho} = \mathcal{M}^{-1} [U^{\dagger} \ket{{b}}\bra{{b}} U]$. 
By construction, this implies that $\mathbb{E}[\hat{\rho}] = \rho$. 
Estimating the energy of any observable, say like the Hamiltonian $H$ from Equation~\ref{eq:hamiltonian}, then follows as $\Tr(H\rho) = \mathbb{E}[\Tr(H\hat{\rho})]$. 
\end{enumerate}


We repeat this process $N$ times in  order to get $N$ independent snapshots from which to compute $\mathbb{E}[\Tr(H\hat{\rho})]$. 
The average here is taken by assuming equal probability for the choice of measurement settings and further averaging over the various measurement outcomes for each setting. 
It is known from Ref.~\cite{Huang2020} that for the family of random Pauli basis measurements on each qubit, the inverse channel for two qubits $\mathcal{M}^{-1} = \mathcal{M}^{-1}_1 \otimes \mathcal{M}^{-1}_2$ where $\mathcal{M}^{-1}_i(A_i) = 3A_i - \mathbb{I}$ is a single-qubit operation. 
Here, $A_i$ denotes the snapshot for the $i^{th}$ qubit, $U^\dagger_i |b_i\rangle \langle b_i| U_i$, where $b_i$ is the measurement outcome and $U_i$ maps a Pauli $P_i$ to the $\pauliZ$ basis. 

\begin{SCfigure}[1.5]
\begin{tikzpicture}[node distance = 1.5cm]
\node (cqcdata) [process] {quantum measurement};
\node (estimator) [data, below of=cqcdata] {construct estimator};
\node (predict) [process, below of=estimator] {estimate chemical property};
\node (properties) [decision, below of=predict, yshift=-0.5cm] {higher accuracy?};
\node (finish) [startstop, below of=properties,yshift=-0.5cm] {finish};
\draw [arrow] (cqcdata) -- (estimator);
\draw [arrow] (estimator) -- (predict);
\draw [arrow] (predict) -- (properties);
\draw [arrow] (properties.west) -- node[near start, above] {yes} ++(-1.5, 0) |- (cqcdata.west);
\draw [arrow] (properties) -- (finish) node[near start, right] {no};
\end{tikzpicture}
\caption{%
Flowchart of how the classical shadows technique is used in the demonstration. 
Starting with the measurement outcomes collected from the quantum experiments, a succinct description of the quantum state is constructed. 
This description is guaranteed to be an unbiased estimator $\hat{\rho}$ for the quantum state $\rho$ prepared with the quantum circuit. 
When chemically relevant properties are given as observables $O_1, O_2, \ldots$, the state's behavior can be predicted classically as $\Tr(\hat{\rho}O_i)$. 
Note that it suffices to construct the estimator just once as it can be reused to predict different properties of the state. More measurement could be conducted if a higher accuracy is desired.
\label{figure:flowchart_shadows}}
\end{SCfigure}
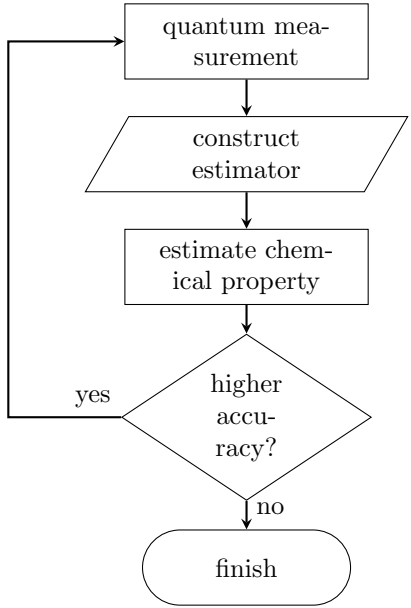

Using the just described method for $N = O(1/\epsilon^2)$ samples, we can construct an implicit, unbiased single-shot estimator for the two-qubit ground state that can provide $\epsilon$-accurate estimates for various chemical properties, such as the ground state energy in our case. 
A flowchart depicting how classical shadows can be used to predict other chemically relevant properties is shown in Figure~\ref{figure:flowchart_shadows}.


\sidecaptionvpos{figure}{c}

We favor this approach over full state tomography or individually estimating the expectation of each Hamiltonian term because the number of samples required for the estimator is independent of system size and depends only logarithmically on the number of chemically relevant properties to be estimated or the number of terms.
As we expand to larger system sizes, this can translate into significant savings in number of samples.
In this case, we take the ground state energy as the main property to train the classical shadows model. 
Additionally, by taking advantage of the symmetries in the ground state of an active space mapped that are known a priori, we can create the classical shadows estimator just by measuring each of the two qubits randomly in the $\pauliX$ or $\pauliZ$ basis.

\section{Encoding the Quantum Computation}
To implement the circuit in Equation~\ref{eq:logical_circuit}, we encode two logical qubits with the  $[[4,2,2]]$ code (denoted as \Cfour)~\cite{vaidman1996,Grassl_1997,C4code}, which is a four-qubit error-detecting code.   
This encoding permits fault-tolerant instantiations of the necessary stabilizer operations---state preparation, CNOT and measurement---up to distance two.  
That is, the code enables detection of single faults in the operations, and rejection of the computation upon such detection, if necessary.
Pauli $\pauliY$ rotations are implemented by teleportation into the encoded \Cfour qubits.  
The circuit implementation of these rotations is not fault tolerant: some faults may go undetected.  
Our circuits attempt to minimize the set of undetectable errors, as explained below.

The circuit template for each experiment is illustrated in Figure~\ref{figure:c4-zz} and it can be partitioned roughly into four parts: state preparation, state verification, $\pauliY$ rotations, and Pauli measurements.
The first three parts of the circuit are fixed, while the Pauli measurements change depending on the chosen measurement bases.

The state preparation uses a variant of an idea by Knill~\cite{knill2004}.
The idea is to first prepare an entangled state that can then be used to teleport a rotation into the codeword.  
In this case, we prepare two Bell pairs. 
Half of each pair is encoded in \Cfour and the other half of each pair is supported on a single physical qubit.
The entire six-qubit stabilizer state has distance two and can be checked for errors.
Two weight-four stabilizers are measured using an interleaving technique by Reichardt~\cite{Reichardt_2020}.  
This circuit detects all weight-two errors generated by a single fault in preparation.

Next, rotations $\RY(\alpha)$ and $\RY(\beta)$ are performed on the two unencoded halves of the Bell pairs.  
Measurement of each qubit then teleports the rotation into the corresponding encoded \Cfour half.  
The effective circuit for each rotation is illustrated in Figure~\ref{figure:Rp-teleportation-bell}. 
To suppress the effect of measurement errors, we first prepare another qubit and ``copy'' the bit value using a CNOT so that the two qubits together form a classical error-detecting code. 
Both qubits are then measured.  
If the measurements agree, then computation proceeds.  
If the measurements disagree, then an error has been detected and the computation is rejected.

The third rotation, $\RY(\gamma)$, is also implemented by teleportation of a physical qubit.  
In this case, however, we do not have the advantage of prior entanglement with the \Cfour qubits.  
Instead, we execute a logical $\pauliI\otimes \pauliY \equiv \pauliI\otimes \pauliX \otimes \pauliZ \otimes \pauliY$ operation on \Cfour, controlled on the physical qubit.  
Subsequent rotation and measurement of the physical qubit then teleports a logical $\RY$ rotation, as depicted in Figure~\ref{figure:Rp-teleportation-arbitrary}.  
Again, measurement error is suppressed by copying the bit and measuring twice.

Finally, the two \Cfour qubits are measured.  
The $\pauliZZ$ and $\pauliXX$ measurements can be executed transversally.  
Mixed basis $\pauliXZ$ and $\pauliZX$ are executed by first non-destructively measuring $\pauliIX$ (or $\pauliXI$) twice. 
Computation is rejected if the two outcomes disagree.

The CNOT in Equation~\ref{eq:logical_circuit} is notably missing from Figure~\ref{figure:c4-zz}.
The \Cfour code admits a CNOT by swapping qubits two and four within the encoded block.  
This swap is accomplished implicitly by relabeling the qubits and adjusting subsequent gates in the circuit, as appropriate.

Each of the $\RY$ teleportations require tracking of Pauli $\pauliY$ frames through the circuit.  
The $\RY(\alpha)$ frame anti-commutes with the subsequent $\RY(\gamma)$ rotation.  
We account for this by accepting only those cases in which the $\RY(\alpha)$ frame is trivial.  
Alternatives to this approach include conditional real-time application of the $\pauliY$ frame, or direct encoding of $\RY(\alpha)|0\rangle$ into \Cfour.
For simplicity, we did not attempt those approaches here.

\begin{figure}
\centering
\begin{subfigure}[b]{\textwidth}
    \centering
\mbox{
\Qcircuit @C=.3em @R=.3em {
\lstick{\ket{0}} &\qw      &\qw     &\qw      &\targ    &\targ    &\qw &\qw             &\qw      &\qw     &\qw      &\qw     &\qw      &\qw     &\qw      &\qw     &\qw &\qw &\qw                     &\qw     &\qw &\qw &\qw &\qw &\qw &\qw &\qw            &\qw       &\qw     &\qw           &\qw                 &\qw     &\qw &\multimeasureD{3}{\text{measure}}\\
\lstick{\ket{+}} &\qw      &\ctrl{3}&\targ    &\qw      &\ctrl{-1}&\qw &\qw             &\qw      &\qw     &\qw      &\qw     &\qw      &\ctrl{8}&\targ    &\qw     &\qw &\qw &\qw                     &\qw     &\qw &\qw &\qw &\qw &\qw &\qw &\qw            &\qw       &\qw     &\gate{\pauliY}&\qw                 &\qw     &\qw &\ghost{\text{measure}}\\
\lstick{\ket{+}} &\ctrl{4} &\qw     &\qw      &\ctrl{-2}&\targ    &\qw &\qw             &\qw      &\qw     &\qw      &\qw     &\targ    &\qw     &\qw      &\ctrl{7}&\qw &\qw &\qw                     &\qw     &\qw &\qw &\qw &\qw &\qw &\qw &\qw            &\qw       &\ctrl{0}&\qw           &\qw                 &\qw     &\qw &\ghost{\text{measure}}\\
\lstick{\ket{+}} &\qw      &\qw     &\ctrl{-2}&\qw      &\ctrl{-1}&\qw &\qw             &\qw      &\qw     &\qw      &\qw     &\qw      &\qw     &\qw      &\qw     &\qw &\qw &\qw                     &\qw     &\qw &\qw &\qw &\qw &\qw &\qw &\qw            &\targ     &\qw     &\qw           &\qw                 &\qw     &\qw &\ghost{\text{measure}}\\
\lstick{\ket{0}} &\qw      &\targ   &\qw      &\qw      &\qw      &\qw &\qw             &\targ    &\qw     &\qw      &\ctrl{5}&\qw      &\qw     &\qw     &\qw      &\qw &\qw &\gate{\RX(\alpha)\gateS}&\ctrl{1}&\measureD{\pauliZ}\\
                 &         &        &         &         &         &    &                &         &        &         &        &         &        &         &        &    &    &\lstick{\ket{0}}        &\targ   &\measureD{\pauliZ}\\
\lstick{\ket{0}} &\targ    &\qw     &\qw      &\qw      &\qw      &\qw &\qw             &\qw      &\ctrl{3}&\targ    &\qw     &\qw      &\qw     &\qw      &\qw     &\qw &\qw &\gate{\RX(\beta)S}      &\ctrl{1}&\measureD{\pauliZ}\\
                 &         &        &         &         &         &    &                &         &        &         &        &         &        &         &        &    &    &\lstick{\ket{0}}        &\targ   &\measureD{\pauliZ}\\
                 &         &        &         &         &         &    &\lstick{\ket{+}}&\ctrl{-4}&\qw     &\ctrl{-2}&\qw     &\ctrl{-6}&\qw     &\ctrl{-7}&\qw     &\measureD{\pauliX}\\
                 &         &        &         &         &         &    &\lstick{\ket{0}}&\qw      &\targ   &\qw      &\targ   &\qw      &\targ   &\qw      &\targ   &\measureD{\pauliZ}\\
                 &         &        &         &         &         &    &                &         &        &         &        &         &        &         &        &    &    &                        &        &    &    &    &    &    &    &\lstick{\ket{+}}&\ctrl{-7}&\ctrl{-8}&\ctrl{-9}    &\gate{\RX{(\gamma)}}&\ctrl{1}&\measureD{\pauliZ}\\
                 &         &        &         &         &         &    &                &         &        &         &        &         &        &         &        &    &    &                        &        &    &    &    &    &    &    &                &         &         &             &\lstick{\ket{0}}    &\targ   &\measureD{\pauliZ}
\gategroup{1}{6}{7}{1}{.5em}{..}
\gategroup{2}{9}{10}{17}{.5em}{..}
\gategroup{5}{19}{6}{21}{.5em}{..}
\gategroup{7}{19}{8}{21}{.5em}{..}
\gategroup{1}{26}{12}{33}{.5em}{..}
}
}
    \caption{Circuit template for the encoded equivalent of logical circuit of Equation~\ref{eq:logical_circuit}, followed by a measurement of either logical $\pauliZZ$, $\pauliXX$, $\pauliXZ$, or $\pauliZX$.}  
    \label{figure:c4-zz}
\end{subfigure}
\begin{subfigure}[b]{.2\textwidth}
    \centering
\mbox{
\Qcircuit @C=1em @R=.1em {
&\qw&\measureD{\pauliZ}\\
&\qw&\measureD{\pauliZ}\\
&\qw&\measureD{\pauliZ}\\
&\qw&\measureD{\pauliZ}\\
}
}
    \caption{$\pauliZZ$ measurement}  
\end{subfigure}
\begin{subfigure}[b]{.2\textwidth}
    \centering
\mbox{
\Qcircuit @C=1em @R=.1em {
&\qw&\measureD{\pauliX}\\
&\qw&\measureD{\pauliX}\\
&\qw&\measureD{\pauliX}\\
&\qw&\measureD{\pauliX}\\
& 
}
}
    \caption{$\pauliXX$ measurement}  
\end{subfigure}
\begin{subfigure}[b]{.29\textwidth}
    \centering
\mbox{
\Qcircuit @C=.5em @R=.1em {
&&&&&&&\push{{2\times}}\\
&\qw&\qw&\qw&\qw&\qw             &\targ    &\qw      &\qw         &\measureD{\pauliZ}\\
&\qw&\qw&\qw&\qw&\qw             &\qw      &\qw      &\qw         &\measureD{\pauliZ}\\
&\qw&\qw&\qw&\qw&\qw             &\qw      &\qw      &\qw         &\measureD{\pauliZ}\\
&\qw&\qw&\qw&\qw&\qw             &\qw      &\targ    &\qw         &\measureD{\pauliZ}\\
&   &   &   &   &\lstick{\ket{+}}&\ctrl{-4}&\ctrl{-1}&\measureD{\pauliX}
\gategroup{2}{3}{6}{9}{.5em}{..}
}
}
    \caption{$\pauliXZ$ measurement}  
\end{subfigure}
\begin{subfigure}[b]{.29\textwidth}
    \centering
\mbox{
\Qcircuit @C=.5em @R=.1em {
&&&&&&&\push{{2\times}}\\
&\qw&\qw&\qw&\qw&\qw             &\qw      &\qw      &\qw         &\measureD{\pauliZ}\\
&\qw&\qw&\qw&\qw&\qw             &\targ    &\qw      &\qw         &\measureD{\pauliZ}\\
&\qw&\qw&\qw&\qw&\qw             &\qw      &\qw      &\qw         &\measureD{\pauliZ}\\
&\qw&\qw&\qw&\qw&\qw             &\qw      &\targ    &\qw         &\measureD{\pauliZ}\\
&   &   &   &   &\lstick{\ket{+}}&\ctrl{-3}&\ctrl{-1}&\measureD{\pauliX}
\gategroup{2}{3}{6}{9}{.5em}{..}
}
}
    \caption{$\pauliZX$ measurement}  
\end{subfigure}
\caption{%
(a) Circuit sub-components are indicated by dashed boxes.
First, two Bell pairs are prepared across six qubits.  Half of each pair is encoded in \Cfour. 
Next, weight-4 stabilizers are measured to detect errors. Pauli-$\pauliY$ rotations are then implemented by teleportation of physical qubits with the help of circuit identities shown in Figures~\ref{figure:Rp-teleportation-bell}~and~\ref{figure:Rp-teleportation-arbitrary}.
Measurement of the \Cfour qubits depend on the chosen measurement bases as illustrated in (b), (c), (d) and (e).  
Projective $\pauliX$-basis measurements in (d) and (e) require an extra qubit and are executed twice in order to suppress measurement error.
}
\end{figure}

\begin{figure}[t]
\begin{subfigure}[t]{.45\textwidth}
    \centering
\mbox{
\Qcircuit @C=1.4em @R=1.2em {
\lstick{\ket{+}} & \ctrl{1} & \gate{\gateS} & \gate{\RX(\theta)} & \measureD{\pauliZ} \cwx[1] & \\
\lstick{\ket{0}} & \targ    & \qw               & \qw       &  \gate{\pauliY}             & \rstick{\RY(\theta)\ket{0}} \qw
}
}
    \caption{%
        Preparation of $\RY(\theta)|0\rangle$ by teleportation.
    }  
    \label{figure:Rp-teleportation-bell}
\end{subfigure}
\hfill
\begin{subfigure}[t]{.45\textwidth}
    \centering
\mbox{
\Qcircuit @C=1.4em @R=1.2em {
\lstick{\ket{+}}    & \ctrl{1} & \gate{\RX(\theta)} & \measureD{\pauliZ} \cwx[1] & \\
\lstick{\ket{\psi}} & \gate{\pauliY} & \qw                & \gate{\pauliY}             & \rstick{\RY(\theta)\ket{\psi}} \qw
}
}
    \caption{%
        Rotation $\RY(\theta)$ by teleportation.
    }  
    \label{figure:Rp-teleportation-arbitrary}
\end{subfigure}
\caption{%
Teleportation identifies used to implement the three $\RY$ rotations of the logical circuit of Equation~\ref{eq:logical_circuit}.}
\end{figure}
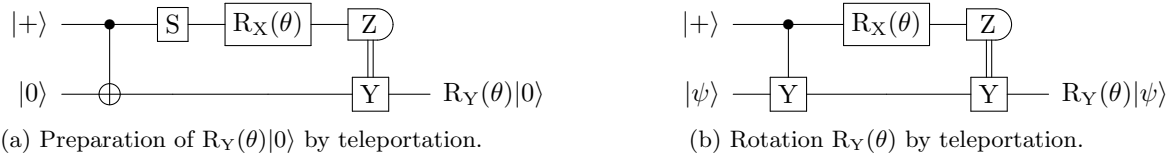

\section{Results and Discussion}
We apply our hybrid workflow to a real chemistry problem.
We start with HPC calculations, using AutoRXN to explore the acetophenone asymmetric hydrogenation reaction, identifying a network of 2227 elementary steps with 3150 unique configurations. Among the 5 configurations that have strong correlation, for demonstration purposes, we select the one that is a transition state of hydrogen loading onto the catalyst (as shown in Figure~\ref{figure:active_space}), and express its ground state wave function using a quantum circuit following  Equation~\ref{eq:logical_circuit} to ~\ref{eq:optimal_angles}.

We prepare both an unencoded version of the quantum circuit, using physical qubits, and an encoded version of the circuit as described in Section \ref{eq:logical_circuit}.
Both circuits were implemented on Quantinuum's H1 ion-trap quantum system.
The unencoded circuit (cf.~Equation~\ref{eq:logical_circuit}) was executed with four different measurement settings $\pauliXX$, $\pauliXZ$, $\pauliZX$, and $\pauliZZ$. 

For the encoded circuit, the logical operations described in Equation~\ref{eq:logical_circuit} were first encoded using the circuit template illustrated in Figure~\ref{figure:c4-zz}. 
The encoded computation used $13$ of the $20$ physical qubits of the H1-system and $66$ physical gates, including $24$ two-qubit gates. 
The $13$ qubits were used in the following manner: 
\begin{itemize}
\item $4$ qubits to encode the two logical qubits using $C_4$ code
\item $3\times 2$ qubits to create and execute the three unencoded non-Clifford rotations
\item $3$ flag qubits for the error detection and measurements
\end{itemize}
This circuit was executed with the four logical measurement settings $\logicalpauliXX$, $\logicalpauliXZ$, $\logicalpauliZX$, and $\logicalpauliZZ$ and the accepted outcomes were decoded into their corresponding logical outcomes. 

For both unencoded and encoded circuits, the measurement outcomes for each circuit were collected and used with the classical shadows technique to construct the implicit estimate $\hat{\rho}$ for the state prepared by the circuit.  
Using this model, we calculated the eight expectations $\expectation{\pauliIX}$, $\expectation{\pauliXI}$, $\expectation{\pauliIZ}$, $\expectation{\pauliZI}$, $\expectation{\pauliXX}$, $\expectation{\pauliXZ}$, $\expectation{\pauliZX}$, $\expectation{\pauliZZ}$.  
Finally, based on these values and the fact that $\hat{E}_\mathrm{gs} = \Tr(\hat{\rho}H)$, the energy of the prepared state was estimated to be 
\begin{align}
\expectation{H} & = 
- 1.99134 
- 0.02882925 (\expectation{\pauliIX}+\expectation{\pauliXI})
+ 0.0541175  (\expectation{\pauliIZ}+\expectation{\pauliZI}) \\
& \quad 
+ 0.01495595 \expectation{\pauliXX}
+ 0.000151287 (\expectation{\pauliXZ}+\expectation{\pauliZX})
+ 0.05900925 \expectation{\pauliZZ}. \nonumber
\end{align}

\begin{figure}
\begin{center}
\includegraphics[width=0.7\textwidth]{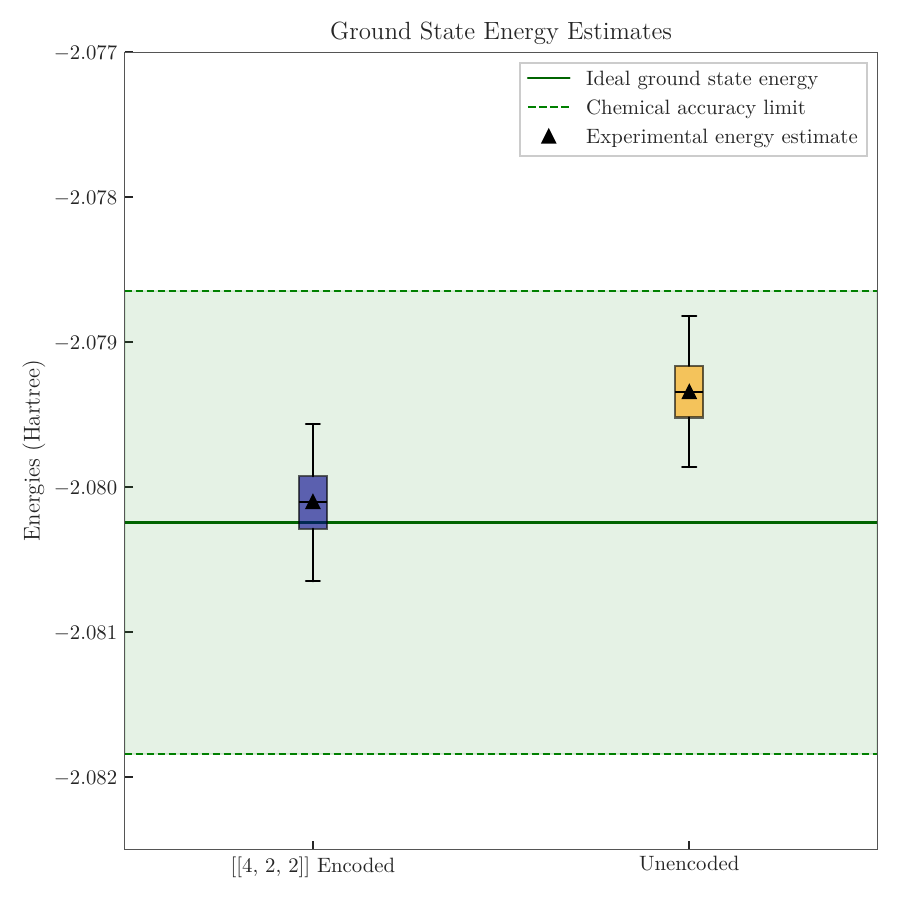} 
\end{center}
\caption{%
Comparison of ground state energy estimated using classical shadows on $[[4,2,2]]$ encoded qubits and unencoded qubits. 
Both data point are presented as a box-and-whisker plot where the interquartile range (box), 95\% confidence intervals (whiskers), and median (line) were determined by statistically bootstrapping the experimental data. 
The mean energy estimation is indicated by a triangle.
The shaded green area indicates a $\pm \qty{1.6}{\milli\hartree}$ error bound following the chemical accuracy definition on the exact ground state energy.
\label{figure:energy_estimates}}
\end{figure}

The encoded computation yielded an estimate of \num{-2.0801(5)} Hartree, while the unencoded computation had an estimate of \num{-2.0793(5)} Hartree.
The true ground state energy equals \qty{-2.08025}{\hartree}.
The unencoded output state was sampled using \num{40000} shots for each of the four measurement settings $\pauliXX$, $\pauliXZ$, $\pauliZX$, $\pauliZZ$, thus yielding a total of \num{160000} shots. 
The encoded output state was sampled using $\num{80000}$ shots for each of the four logical measurement settings.
Overall, these circuits had a combined acceptance rate of about $47\%$ and the measurement outcomes from these circuits were collected by post-selecting on the slightly less than $\num{150000}$ accepted shots.
Error detection yielded a $3\%$ rejection rate and the use of teleportation flags was responsible for the additional $50\%$ rejection rate. 
We note that this use of teleportation flags can be eliminated by using slightly larger circuits.


To determine the error estimates, we used bootstrap techniques and resampled the data sets \num{5000} times to calculate the $95\%$ confidence intervals and the interquartile ranges of the two estimates. 
As can be seen by the whiskers in Figure~\ref{figure:energy_estimates}, the encoded computation sampled the true ground state energy within its $95\%$ confidence interval, while using the corresponding unencoded quantum computation, the $95\%$ confidence interval deviates from the true ground state energy. 
This demonstrates the advantage of the encoded computation, and how reliable quantum computation with logical qubits will be extremely beneficial for larger active spaces where much more measurements are needed and deep quantum computations will be required, with very low logical error rates.
 
\begin{figure}
    \centering
    \includegraphics[width=0.9\textwidth]{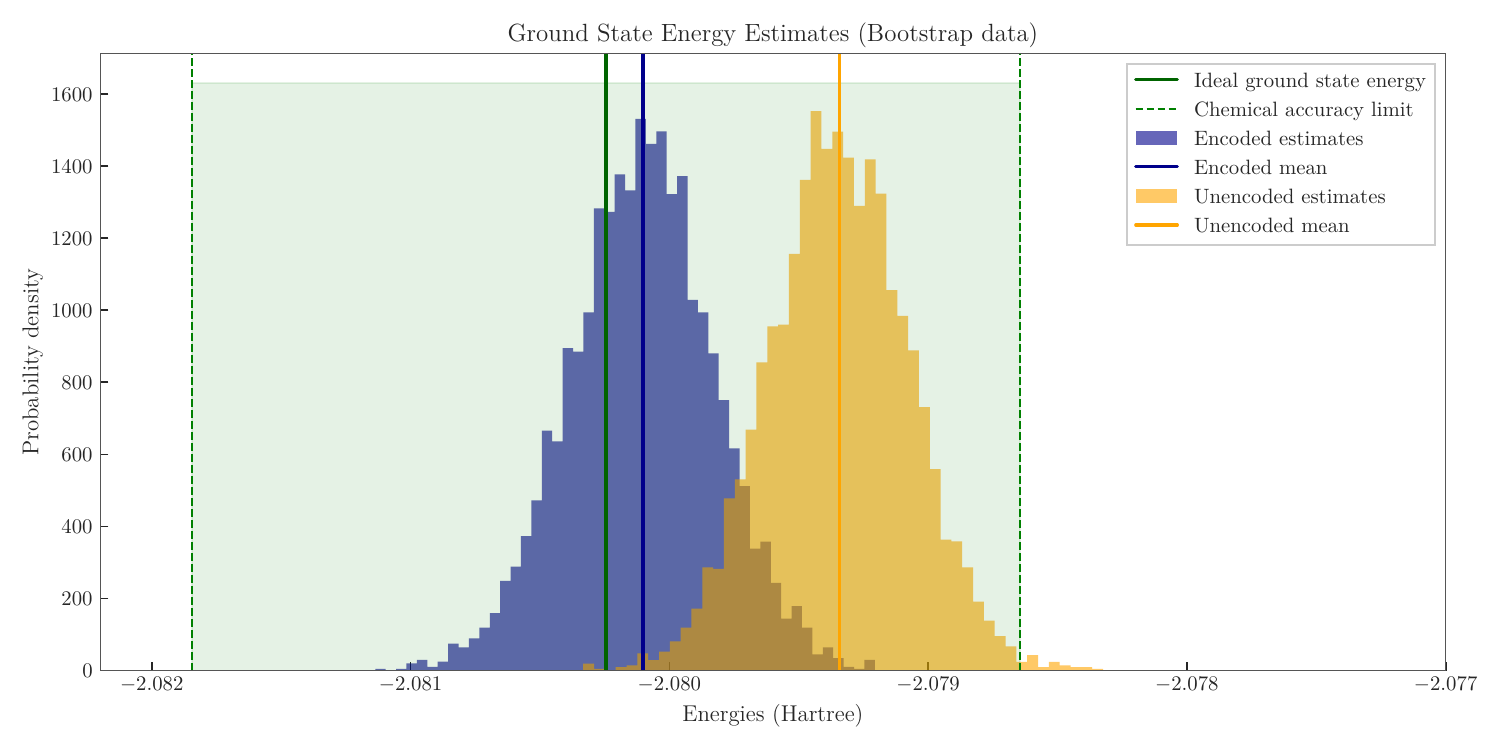}
    \caption{%
    Using \num{5000} bootstrapping samples of the original data gives the two histograms shown here. 
    From this data it can be calculated that with probability greater than $97\%$ the encoded circuit estimation will be better than the unencoded estimation. 
    The full sets of \num{5000} samples go beyond the calculated $95\%$ confidence intervals, which explains why these histograms have a wider range than the whiskers of Figure~\ref{figure:energy_estimates}. 
    \label{fig:failure_prob}}
\end{figure}

The boxes in the same figure also show that the $50\%$ confidence intervals do not overlap, suggesting that the encoded quantum computation performed better than the unencoded computation. 
Using the same bootstrapping data we compared the distributions of the two estimates, as shown in Figure~\ref{fig:failure_prob}. 
This comparison indeed confirms that with $97\%$ likelihood the encoded computation yielded a better estimate for the ground state energy than the unencoded computation. 

\begin{figure}
\begin{center}
\includegraphics[width=0.9\textwidth]{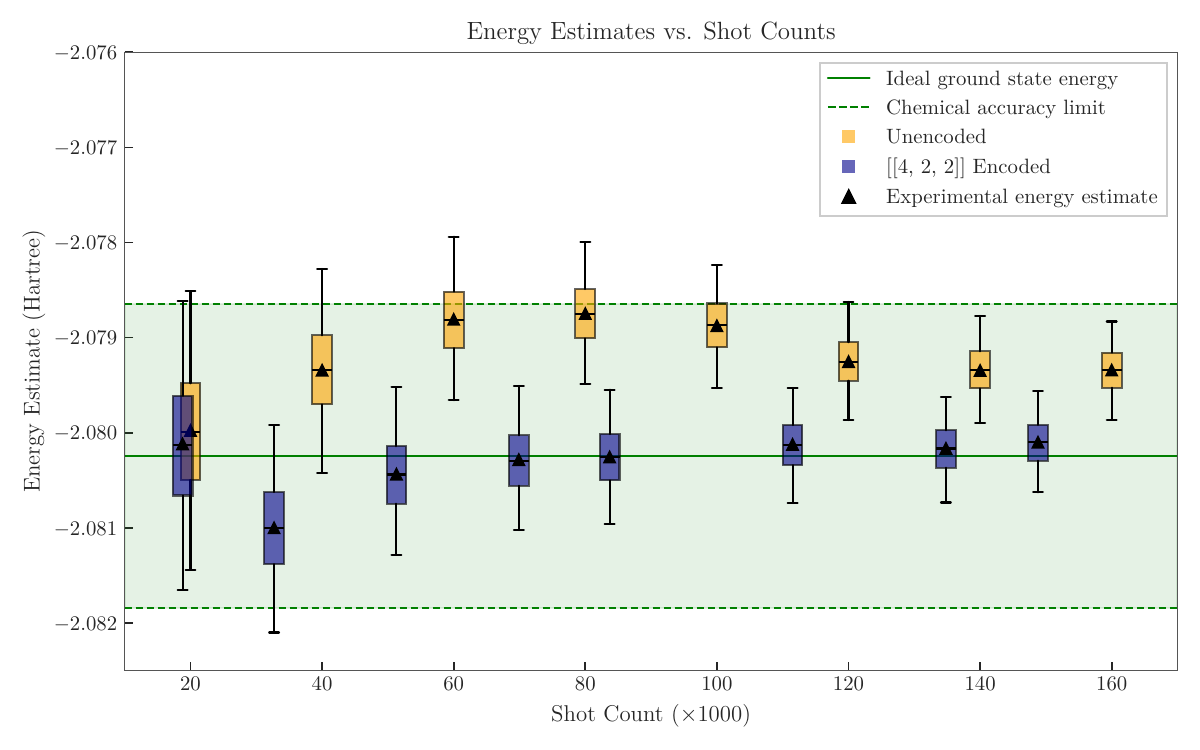} 
\end{center}
\caption{%
The trends of how the energy estimates vary with an increasing number of shots for both the unencoded (light) and $[[4, 2, 2]]$ encoded (dark) circuits. 
The data points are in increments of approximately \num{20000} shots for each circuit type. 
The data for the encoded circuit is not at exact 20k shot intervals due to post-selection of only accepted shots. 
Each data point is presented as a box-and-whisker plot where the interquartile range (box), 95\% confidence intervals (whiskers) and median (orange line) were determined by statistically bootstrapping the experimental data. 
The energy estimated from experimental data is plotted with black triangles.
\label{figure:estimates_vs_shots}}
\end{figure}

How the estimates improved as a function of the number of shots is shown in Figure~\ref{figure:estimates_vs_shots}. 
As can be seen, the encoded circuit consistently samples the true ground state energy within its confidence interval, while the physical circuits cannot improve the sampling accuracy with increased number of shots.
 
\section{Outlook}

We have presented the first end-to-end workflow and demonstration using a combination of cloud HPC, reliable quantum computing, and AI to solve a real chemistry problem.
We compare using both unencoded and encoded quantum circuits in the quantum computational component of our workflow, in both cases preparing the ground state wave function on two qubits (physical or logical, respectively).
We demonstrate a proof-of-concept hybrid workflow, and discuss how aspects of the workflow will scale for larger active site configurations.  We additionally present evidence of the encoded quantum computation yielding results that perform better than when using the unencoded, NISQ computation.

An obvious step forward is to replace the preparation of a classically computed state by an optimization of a wave function using, for example, the VQE algorithm. 
When advancing to larger active spaces, the number of parameters may also grow exponentially in the VQE framework, therefore, algorithms like QPE will be further advantageous to propagate the true eigenstate. At that stage, it will be essential to run QPE with reliable, logical qubits to enable successful deeper quantum computation.
Another direction, combining with the classical shadows technique, is to estimate more properties for the molecules of interest, for example polarizabilities, or spin correlation functions, just to name a few. 
In summary, the workflow we present can be generally applicable to most (homogeneous) reaction problems, and is the first step on the journey to achieve practical quantum advantages in real-world chemistry applications. 

\section{Acknowledgements}

We thank the H1-1 hardware team at Quantinuum for making these experiments possible and Simon McAdams and Travis Humble for feedback on an earlier version of this paper.

\bibliographystyle{alpha}
\bibliography{main}

\end{document}